\documentclass[epj]{webofc}
\usepackage[utf8]{inputenc}
\usepackage[varg]{txfonts}   
\usepackage{booktabs}
\usepackage{xcolor}
\definecolor{darkred}{rgb}{0.4,0.0,0.0}
\definecolor{darkgreen}{rgb}{0.0,0.4,0.0}
\definecolor{darkblue}{rgb}{0.0,0.0,0.4}
\usepackage[bookmarks,linktocpage,colorlinks,
    linkcolor = darkred,
    urlcolor  = darkblue,
    citecolor = darkgreen]{hyperref}
\usepackage{bigints}
\wocname{EPJ Web of Conferences}
\woctitle{Lattice2017}
\newcommand{\ei}[1]{\operatorname{Ei}\left(#1\right)}
\newcommand{\erf}[1]{\operatorname{erf}\left(#1\right)}
\begin{document}
%
\selectlanguage{english}
\title{%
Finite continuum quasi distributions from lattice QCD
}
\author{%
\firstname{Christopher} \lastname{Monahan}\inst{1,2}\fnsep\thanks{Speaker, \email{cjm373@uw.edu}}\and
\firstname{Kostas} \lastname{Orginos}\inst{3,4}
}
\institute{%
Institute for Nuclear Theory, University of Washington,
Seattle, WA 98195-1550, USA
\and
New High Energy Theory Center and Department of Physics and 
Astronomy, Rutgers, the State University of New Jersey,
136 Frelinghuysen Road, Piscataway, New Jersey 08854-8019, USA
\and
Physics Department, College of William and Mary,
Williamsburg, Virginia 23187, USA
\and
Thomas Jefferson National Accelerator Facility, Newport News, 
Virginia 23606, USA
}
\abstract{%
We present a new approach to extracting continuum quasi distributions from lattice QCD. Quasi distributions are defined by matrix elements of a Wilson-line operator extended in a spatial direction, evaluated between nucleon states at finite momentum. We propose smearing this extended operator with the gradient flow to render the corresponding matrix elements finite in the continuum limit. This procedure provides a nonperturbative method to remove the power-divergence associated with the Wilson line and the resulting matrix elements can be directly matched to light-front distributions via perturbation theory.
}
\maketitle
\section{Introduction}\label{intro}

Protons and neutrons---nucleons---are the basic, observable building blocks of the visible matter in the Universe. But nucleons are not simple building blocks: they are strongly-coupled, dynamical systems, composed of quarks and gluons bound together by the strong nuclear force. Quantum chromodynamics (QCD) is the gauge theory of the strong nuclear force that, in principle, connects protons and neutrons to their constituent quarks and gluons. In practice, however, piecing together nucleon structure directly from QCD has proved challenging, largely due to the nonperturbative nature of QCD at energy scales around the nucleon mass.

Parton distribution functions (PDFs), which characterise the distribution of a fast-moving nucleon's longitudinal momentum amongst its constituents, have posed particular difficulties for first principles' QCD calculations. PDFs are most naturally formulated as matrix elements of null-separated quark fields, which cannot be calculated using Euclidean lattice QCD, the only ab initio, systematically-improvable method currently available for nonperturbative QCD calculations. In \cite{Ji:2013dva}, Ji proposed a promising method to extract PDFs, from lattice QCD calculations of quasi distributions, which are spatially-extended operators between nucleon states at finite momentum. Related frameworks have also been proposed: in \cite{Ma:2014jla,Ma:2014jga,Ma:2017pxb} quasi distributions were treated as ``lattice cross-sections'' from which light-front PDFs can be factorized and Refs.~\cite{Radyushkin:2016hsy,Orginos:2017kos} introduced and studied the closely-related pseudo distributions.

Quasi and pseudo distributions are matrix elements of time-local operators that can be computed using lattice QCD \cite{Carlson:2017gpk,Briceno:2017cpo} and preliminary nonperturbative results are encouraging \cite{Lin:2014zya,Alexandrou:2015rja,Chen:2016utp,Orginos:2017kos,Zhang:2017bzy,Alexandrou:2017qpu}. Moreover, a number of theoretical issues \cite{Li:2016amo,Carlson:2017gpk,Rossi:2017muf} have now been clarified: the multiplicative renormalisation of the spatial Wilson line operator has been proven \cite{Ji:2017oey,Ishikawa:2017faj}; the factorization of light-front PDFs from quasi distributions was demonstrated in \cite{Ma:2014jla,Ma:2014jga}; and a proof that the matrix element extracted from Euclidean correlation functions is identical to that determined from a Lehmann-Symanzik-Zimmermann reduction procedure in Minkowski spacetime provided in \cite{Briceno:2017cpo} and studied further in \cite{Xiong:2017jtn,Ji:2017rah}. 

On the lattice, there is a power-divergence, which scales exponentially with the length of the Wilson line divided by the lattice spacing, associated with the extended operator that defines quasi and pseudo distributions. Several approaches have been proposed: inspired by heavy quark effective theory, the authors of \cite{Chen:2016fxx,Ishikawa:2016znu} suggested an exponentiated mass renormalisation to remove the power-divergence and, more recently, RI/MOM \cite{Chen:2017mzz,Stewart:2017tvs} and RI$^\prime$ \cite{Alexandrou:2017huk} schemes have been introduced. In \cite{Monahan:2016bvm}, we presented an alternative approach, the smeared quasi distribution, that avoids some of the challenges of the power divergence by taking advantage of the properties of the gradient flow \cite{Narayanan:2006rf,Luscher:2011bx,Luscher:2013cpa}.

The gradient flow exponentially suppresses ultraviolet (UV) field fluctuations, which corresponds to smearing out the original degrees of freedom in real space. Most importantly, the gradient flow renders correlation functions finite \cite{Luscher:2011bx}, up to a multiplicative fermion wavefunction renormalisation \cite{Luscher:2013cpa} and provided one has fixed the renormalised parameters of the original theory. Fixing the flow time, which corresponds to a real-space smearing scale, in physical units, ensures that matrix elements determined at finite lattice spacing remain finite in the continuum limit. The gradient flow therefore enables us to extract finite quasi distributions from lattice calculations \cite{Monahan:2016bvm}. The resulting continuum matrix element can be directly matched to the light-front PDF, or to the quasi distribution renormalised in, for example, the $\overline{MS}$ scheme \cite{Xiong:2013bka,Ji:2015jwa,Constantinou:2017sej,Wang:2017qyg}.

Here we introduce smeared quasi distributions and study the corresponding matrix element at one loop in perturbation theory. We compute the matrix element of the smeared Wilson-line operator between external, gauge-fixed quark states. As we demonstrate, the Wilson-line power divergence appears at one loop as a contribution linear in $\overline{z} = z/r_\tau$, where $z$ is the length of the Wilson line and $r_\tau$ is the gradient flow smearing radius, for $\overline{z} \gg 1$. Subtracting this contribution, the remaining matrix element is finite, and has a well-defined $\overline{z}\to0$ limit, in contrast to the $\overline{MS}$ scheme. In the small flow-time regime, $\overline{z} \gg 1$, the matrix element depends only logarithmically on $\overline{z}$ and thus satisfies a relation similar to a standard renormalisation group equation.

\section{\label{sec:defns}Light-front and quasi distributions}

We consider only flavour nonsinglet unpolarised quasi and light-front PDFs, for which we can ignore mixing with the gluon distribution. Extending our discussion to the polarised distribution is straightforward.

\subsection{Light-front PDFs}

We write the renormalised light-front PDFs as $f(\xi,\mu)$, where we have introduced light-front coordinates $x^\pm=(t\pm z)/\sqrt{2}$, a renormalisation scale $\mu$, and $\xi=k^+/P^+$. In general, we can relate the renormalised light-front PDFs to the bare PDFs, $f^{(0)}(\xi)$, through
\begin{equation}
f(\xi,\mu) = \int_\xi^1 \frac{\mathrm{d}\zeta}{\zeta}{\cal 
Z}\left(\frac{\xi}{\zeta},\mu\right)f^{(0)}(\zeta),
\end{equation}
where the bare PDF is given by \cite{Collins:2011zzd}
\begin{equation}
f^{(0)}(\xi) = \int_{-\infty}^\infty 
\frac{\mathrm{d}\omega^-}{4\pi}
e^{-i\xi P^+\omega^-} \left\langle P \left| 
T\,\overline{\psi}(0,\omega^-,\mathbf{0}_\mathrm{T})
W(\omega^-,0)\gamma^+\frac{\lambda^a}{2}\psi(0) \right| 
P\right\rangle_\mathrm{C}.
\end{equation}
Here $T$ indicates time-ordering, $\psi$ is a quark 
field, and we include only connected 
contributions, represented by the subscript $C$. The Wilson line operator is
\begin{equation}
W(\omega^-,0) = 
{\cal P}\exp\left[-ig_0\int_0^{\omega^-}\mathrm{d}y^-A^+_a(0,y^-,
\mathbf{0}_{\mathrm{T}})T_a\right],
\end{equation}
where ${\cal P}$ is the path-ordering operator, $g_0$ the QCD bare coupling, 
and $A_\alpha = A_\alpha^a T^a$ is the $SU(3)$ gauge potential with generator 
$T^a$ (summation over the color index $a$ implicit). The nucleon state, $|P\rangle$, is a spin-averaged, exact momentum eigenstate with 
relativistic normalisation
\begin{equation}
\langle P' | P \rangle = 
(2\pi)^32P^+\delta\left(P^+-P^{\prime\,+}\right)\delta^{(2)}
\left(\mathbf{P}_\mathrm{T} - \mathbf{P}'_\mathrm{T}\right).
\end{equation}

The renormalised nonsinglet PDFs satisfy a DGLAP equation 
\cite{Gribov:1972rt,Dokshitzer:1977sg,Altarelli:1977zs} that describes their 
scale dependence
\begin{equation}
\mu\frac{\mathrm{d}\, f(x,\mu)}{\mathrm{d}\mu} = \frac{\alpha_s(\mu)}{\pi} 
\bigintsss_x^1 
\frac{\mathrm{d}y}{y} f(y,\mu) P\left(\frac{x}{y}\right),
\end{equation}
were $P\left(z\right)$ is a function whose moments are given by
\begin{equation}\label{eq:gamman}
\int_0^1 \mathrm{d}x\, x^{n-1} P(x) = \gamma^{(n)}.
\end{equation}
The anomalous dimensions, $\gamma^{(n)}$, satisfy
\begin{equation}\label{eq:rgan}
\left[\mu\frac{\mathrm{d}\;}{\mathrm{d}\mu} -\frac{\alpha_s(\mu)}{\pi}  
\gamma^{(n)} \right]a^{(n)}(\mu) = 0.
\end{equation}
Here $\alpha_s(\mu)$ is the (renormalised) strong coupling constant and the $a^{(n)}(\mu)$ are the renormalised Mellin 
moments of the PDF,
\begin{equation}
a^{(n)}(\mu) = \int_0^1 \mathrm{d}\xi\,\xi^{n-1}\left[
f(\xi,\mu)+(-1)^n\overline{f}(\xi,\mu)\right] = \int_{-1}^1 
\mathrm{d}\xi\,\xi^{n-1}
f(\xi,\mu),
\end{equation}
which are related to matrix elements of renormalised twist-two operators, 
${\cal O}^{\{\nu_1\ldots \nu_n\}}(\mu) =Z_{\cal O}(\mu){\cal 
O}_0^{\{\nu_1\ldots \nu_n\}}$, through
\begin{equation}\label{eq:RopePDF}
\left\langle P | {\cal O}^{\{\nu_1\ldots \nu_n\}}(\mu) | P 
\right\rangle = 2a^{(n)}(\mu) \left(P^{\nu_1}\cdots P^{\nu_n} - 
\mathrm{traces}\right).
\end{equation}

\subsection{Smeared quasi distributions}

In \cite{Monahan:2016bvm}, we proposed a new approach to extracting continuum quasi distributions from lattice calculations, constructing finite matrix elements by smearing both the fermion and gauge fields via the gradient flow 
\cite{Narayanan:2006rf,Luscher:2011bx,Luscher:2013cpa}. Further details of the gradient flow, and a variety of applications in lattice calculations, are given in the reviews 
\cite{Luscher:2013vga,Ramos:2015dla}. 

We use ringed fermion fields at flow time $\tau$ \cite{Makino:2014taa,Hieda:2016lly}, denoted by $\overline{\chi}(x;\tau)$ and $\chi(x;\tau)$, and construct a smeared Wilson line operator, ${\bf\cal W}(x_1,x_2;\tau)$, from smeared gauge fields $B_\alpha(x;\tau)$. We then define the connected matrix element
\begin{equation}\label{eq:hdef}
h_\alpha^{(s)}\left(\frac{n^2}{\tau},n\cdot P,
\sqrt{\tau}\Lambda_{\mathrm{QCD}},\sqrt{\tau}M_\mathrm{N}\right) =
\frac{1}{2P_z}\left\langle 
P_z \left| \overline{\chi}(n;\tau) 
{\bf\cal W}(0,n;\tau)\gamma_\alpha\frac{\lambda^a}{2}\chi(0;\tau)\right| 
P_z\right\rangle_\mathrm{C},
\end{equation}
which depends only on dimensionless combinations of scales and Euclidean SO(4) invariants \cite{Radyushkin:2016hsy}. The spacetime position of the antiquark field, $n = (\mathbf{0},z,0)$, is a four-vector usually chosen to be in the $z$-direction. Ringed fermion fields require no wavefunction renormalisation and this smeared matrix element remains finite provided the flow time, $\tau > 0$, is fixed in physical units, 
because correlation functions constructed from smeared fields are finite \cite{Luscher:2011bx,Luscher:2013cpa}. As we illustrate later at one loop in perturbation theory, divergences appear in the limit of vanishing flow time and the matrix element requires renormalisation in this limit. The Lorentz index $\alpha$ is arbitrary, although choosing $\alpha = 4$ simplifies lattice calculations by removing mixing at finite lattice spacing \cite{Constantinou:2017sej} and eliminating some higher-twist contamination \cite{Radyushkin:2016hsy}.

We define the smeared quasi distribution \cite{Monahan:2016bvm} as
\begin{equation}\label{eq:qpdfdef}
q^{\,(s)}\left(\xi,\sqrt{\tau}P_z,\sqrt{\tau}\Lambda_{\mathrm{QCD}}, 
\sqrt{\tau}M_\mathrm{N}\right) 
= 
\int_{-\infty}^\infty \frac{\mathrm{d}z}{2\pi} e^{i\xi z 
P_z} P_z h^{(s)}\left(\frac{n^2}{\tau},n\cdot P,\sqrt{\tau}\Lambda_{\mathrm{QCD}}, 
\sqrt{\tau}M_\mathrm{N}\right),
\end{equation}
where $\xi$ is best interpreted as a dimensionless momentum variable in a Fourier transformation. Working in the regime in which
\begin{equation}\label{eq:scales_tau}
\Lambda_{QCD},M_N\ll P_z \ll \tau^{-1/2},
\end{equation}
the smeared quasi distribution can be directly related to the light-front distribution by \cite{Monahan:2016bvm}
\begin{equation}
q^{\,(s)}\left(x,\sqrt{\tau} \Lambda_{\mathrm{QCD}}, \sqrt{\tau}P_z\right) = \int_{-1}^{1} \frac{d\xi}{\xi}\, \widetilde{Z}\left(\frac{x}{\xi},\sqrt{\tau}\mu, 
\sqrt{\tau}P_z\right) 
f(\xi,\mu) + 
{\cal O}(\sqrt{\tau}\Lambda_{\mathrm{QCD}}).
\end{equation}
In analogy to the light-front PDFs, in \cite{Monahan:2016bvm} we derived a DGLAP-like equation for the 
matching kernel that relates smeared quasi PDFs and light-front PDFs, given by
\begin{equation}
 \mu\frac{d\;}{d\mu}\,\widetilde{Z}\left(x,\sqrt{\tau}\mu, \sqrt{\tau}P_z\right) =  -
\frac{\alpha_s(\mu)}{\pi} \bigintsss_{x}^\infty \frac{dy}{y}  
\widetilde{Z}\left(y,\sqrt{\tau}\mu, \sqrt{\tau}P_z\right) P\left(\frac{x}{y}\right),
\end{equation}
up to corrections of ${\cal O}(\sqrt{\tau}\Lambda_{\mathrm{QCD}})$.

The quasi distribution and the light-front PDF have the same infrared (IR) structure \cite{Carlson:2017gpk,Briceno:2017cpo}, so that the matching kernel can be determined in perturbation theory. In this work, we focus on the one-loop renormalisation of the smeared quasi distribution, which can be used to relate the smeared quasi distribution to the quasi distribution at zero flow time. We do not consider the matching between the unsmeared quasi distribution and the light-front PDF, which has been discussed in some detail elsewhere \cite{Xiong:2013bka,Ji:2015jwa}.

\section{Smeared quasi distributions in perturbation theory}

\begin{figure}[thb] 
\centering
\includegraphics[width =0.9\textwidth,clip]{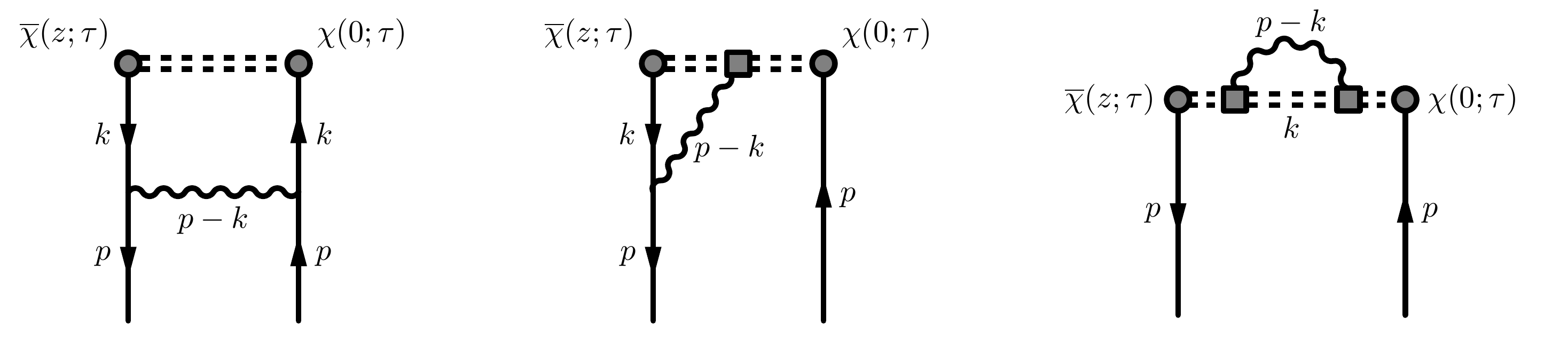}\\
\hspace*{-15pt}(a)\hspace*{0.28\textwidth}(b)\hspace*{0.28\textwidth}(c)\vspace*{10pt}\\
\hspace*{-15pt}\includegraphics[width =0.85\textwidth]{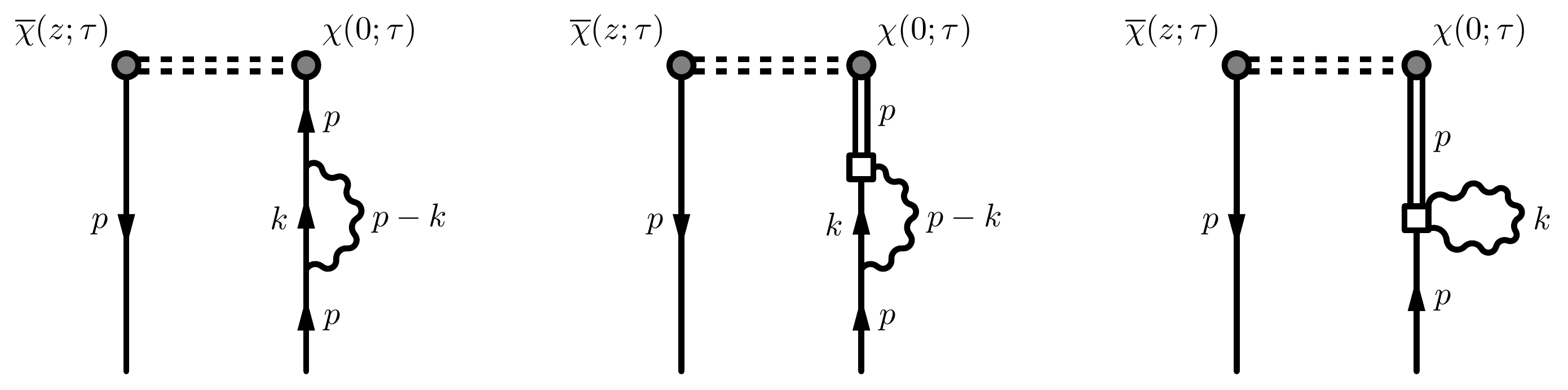}\\
\hspace*{-15pt}(d)\hspace*{0.28\textwidth}(e)\hspace*{0.28\textwidth}(f)\vspace*{10pt}\\
\caption{Diagrams representing the one-loop contributions to the smeared nonsinglet quasi distribution. The grey circles indicate fermion fields at finite flow time $\tau$; solid lines represent fermion propagators; double lines are fermion flow kernels; and open squares are flow vertices at arbitrary flow time $\tau_1$.}
\label{fig:sqpdf1}
\end{figure}

The diagrams in Fig.~\ref{fig:sqpdf1} represent the one-loop contributions to the smeared quasi distribution in generalised Feynman gauge (with $\alpha = \lambda = 1$, where $\alpha$ is the standard gauge-fixing parameter and $\lambda$ is a generalised gauge-fixing parameter at nonzero flow time \cite{Luscher:2010iy,Luscher:2011bx}). Axial gauge simplifies the calculation of the quasi distribution, but cannot be consistently generalised to arbitrary flow time.
We analytically determine the renormalisation of one-loop contributions to the matrix element in Eq.~\eqref{eq:hdef} by calculating the diagrams in Fig.~\ref{fig:sqpdf1} with external quark states at zero momentum. Some of these contributions have IR divergences, which we regulate with dimensional regularisation. We show that, at one-loop in perturbation theory, these IR divergences are independent of the flow time.

Diagram (d) is the usual wavefunction renormalisation, $Z_\psi$, which is independent of the flow time and known well beyond one loop \cite{Chetyrkin:1999pq}; we do not consider this diagram any further here. The other contributions are given by \cite{Monahan:2017hpu}
\begin{align}
h_4^\mathrm{(a)}(\overline{z}^2,\mu^2z^2) = {} & \left(\frac{g_0}{4\pi}\right)^2C_F\gamma_4 \bigg[\frac{1}{\epsilon_{\mathrm{IR}}}-\gamma_{\mathrm{E}}-\log\left(\pi \mu^2z^2\right) +1-\frac{1}{\overline{z}^4}\Big(1-e^{-\overline{z}^2}(1+\overline{z}^2)\Big)+\ei{-\overline{z}^2} \bigg] , \label{eq:hta} \\
h_4^\mathrm{(b)}(\overline{z}^2)  = {} & \left(\frac{g_0}{4\pi}\right)^2C_F \gamma_4 \cdot2\bigg[1-\gamma_{\mathrm{E}}-\log\left(\overline{z}^2\right)+\ei{-\overline{z}^2} + \frac{1}{\overline{z}^2}\left(e^{-\overline{z}^2}-1\right)\bigg], \label{eq:htb}\\
h_4^\mathrm{(c)}(\overline{z}^2) = {} & \left(\frac{g_0}{4\pi}\right)^2C_F \gamma_4 \,2\bigg[\gamma_{\mathrm{E}}+\log\left(\overline{z}^2\right)- \ei{-\overline{z}^2}  - 2\left(e^{-\overline{z}^2}-1\right) -\sqrt{\pi\overline{z}^2}\erf{\sqrt{\overline{z}^2}}\bigg], \label{eq:htc} \\
h_4^\mathrm{(e)}(\mu^2\tau) = {} & \left(\frac{g_0}{4\pi}\right)^2  C_F \gamma_4\cdot 2\left[\frac{1}{\epsilon_{\mathrm{UV}}} 
+1+\log\big(8\pi\mu^2\tau\big)\right], \label{eq:hte}\\
h_4^\mathrm{(f)} (\mu^2\tau) = {} &  -\left(\frac{g_0}{4\pi}\right)^2C_F \gamma_4 \cdot 4\left[\frac{1}{\epsilon_{\mathrm{UV}}} +\frac{1}{2} + \log(8\pi\mu^2\tau)\right] \label{eq:htf}.
\end{align}
Here $C_F = 4/3$ is the color factor; $g_0$ is the bare coupling; $\ei{z}$ is the exponential integral
\begin{equation}
\ei{z} = -\int_{-z}^\infty \frac{e^{-t}}{t}\mathrm{d}t;
\end{equation}
and $\erf{z}$ is the error function
\begin{equation}
\erf{z} = \frac{2}{\sqrt{\pi}}\int_0^z e^{-t^2}\mathrm{d}t.
\end{equation}
Our results for diagrams (e) and (f), which have UV divergences removed by $Z_\chi$, are in agreement with those presented in \cite{Endo:2015iea}

The complete one-loop smeared matrix element is therefore
\begin{equation}
h_4^{(\mathrm{s})}(\overline{z}^2)={\cal Z}^{(4)}(\overline{z}^2)h_4^{(0)} = Z^{(4)}(\overline{z}^2,\mu^2\tau) Z_\chi(\mu^2\tau) Z_\psi^{-1}h_4^{(0)}+{\cal O}(\alpha_s^2),\quad \mathrm{with}\quad h_4^{(0)} = \overline{u}^s(p)\frac{\lambda^a}{2}\gamma_4 u^s(p)
\end{equation}
and 
\begin{equation}
Z^{(4)}(\overline{z}^2,\mu^2\tau) = 1 + \frac{\alpha_s(\mu)}{3\pi}\Big[h_4^\mathrm{(a)}(\overline{z}^2,\mu^2\tau)+h_4^\mathrm{(b)}(\overline{z}^2)  +h_4^\mathrm{(c)}(\overline{z}^2)+h_4^\mathrm{(d)}+h_4^\mathrm{(e)}(\mu^2\tau)+h_4^\mathrm{(f)}(\mu^2\tau)\Big].
\end{equation}
The individual nonzero contributions $h^\mathrm{(i)}$ are listed in Eqs.~\eqref{eq:hta} to \eqref{eq:htf}. Here $\alpha_s(\mu) = g^2/(4\pi)$ is the renormalised coupling constant,  equal to the bare coupling constant at this order, and is most naturally evaluated at the scale $\mu^2 = 1/(8\tau)$.

\subsection{Asymptotic behavior}
The small quark-separation limit, $\overline{z} \to 0$, of these results
\begin{equation}
\lim_{\overline{z} \to 0}h_4^{(s)} = \left(\frac{g_0}{4\pi}\right)^2C_F \gamma_\alpha \left\{\frac{1}{2}-3\left[\frac{1}{\epsilon_{\mathrm{UV}}} + \log\left(8\pi\mu^2\tau\right)\right] \right\},\label{eq:hoztozero}
\end{equation}
in agreement with the local vector-current result of \cite{Hieda:2016lly}. Incorporating $Z_\chi(\mu^2\tau)$ \cite{Luscher:2013cpa,Makino:2014taa}, this leads to
\begin{equation}\label{eq:veclimit}
V_4^{(\mathrm{s})} = \lim_{\overline{z} \to 0}h_4^{(\mathrm{s})}(\overline{z}^2) = \bigg\{ 1 + \frac{\alpha_s}{3\pi}\bigg[\frac{1}{2}-\log(432) \bigg]\bigg\}\overline{\psi}\gamma_4\psi.
\end{equation}
This local vector-current result is finite, as it must be for a conserved vector current, but nonzero, which is typical for composite operators of ringed fermions (see, for example, the nonsinglet axial-vector current of \cite{Endo:2015iea}).

In the small flow-time limit, it is possible to show that these results reduce to the corresponding contributions at zero flow time, in the $\overline{MS}$ scheme \cite{Monahan:2017hpu}. The renormalisation parameter satisfies
\begin{equation}\label{eq:zsmalltau}
{\cal Z}_{\mathrm{sub}}^{(4)}(\overline{z}^2) \stackrel{\overline{z}\gg1}{\simeq} 7 - \gamma_{\mathrm{E}} - \log(432) - \log(\overline{z}^2),
\end{equation}
where the subscript ``sub'' indicates that we have subtracted the one-loop contribution to the power divergence.
In this limit, the parameter ${\cal Z}_{\mathrm{sub}}^{(4)}(\overline{z}^2)$ depends only logarithmically on $\overline{z}$, and therefore satisfies a relation analogous to a typical renormalisation group equation \cite{Monahan:2015lha}:
\begin{equation}\label{eq:zsubRG}
\left[\frac{\mathrm{d}}{\mathrm{d}\log\overline{z}^2} +\gamma_{\overline{z}}\right]{\cal Z}_{\mathrm{sub}}^{(4)}(\overline{z}^2) = 0.
\end{equation}

\section{\label{sec:concs}Conclusion}

PDFs directly relate nucleon structure to the fundamental degrees of freedom of QCD, quarks and gluons. PDFs are defined as matrix 
elements of light-front wave functions, which cannot be directly calculated in 
Euclidean lattice QCD. Although the Mellin moments of PDFs can be 
calculated in lattice QCD, through matrix elements of twist-two operators, these calculations are limited to the first few moments by power-divergent mixing on the lattice.

A new approach to determining PDFs in lattice QCD was recently proposed by Ji, in which one calculates quasi distributions---Fourier transforms of Euclidean matrix elements of spatially-separated quarks fields at large nucleon momentum. Light-front PDFs can then be extracted from these quasi distributions through a suitable matching procedure, using LaMET. Related frameworks, including extracting light-front PDFs from ``lattice cross-sections'', of which quasi distributions are one example, and pseudo distributions, have also been proposed.

Recent lattice calculations have provided promising results, for both quasi and pseudo distributions, but several aspects of these approaches are yet to be 
fully understood. First, there is the practical issue of the systematic 
uncertainties associated with finite nucleon momenta in lattice calculations. 
This issue is likely to be resolved by algorithmic and hardware improvements, to the extent that lattice 
calculations will provide useful input into global analyses where experimental data are inadequate. Second, the 
renormalisation of the extended operator that defines the quasi PDFs is challenging, because of the presence of a power divergence generated by the Wilson line on the lattice.

We proposed one approach to removing this power divergence by introducing a smeared quasi distribution,
constructed from fields smeared via the gradient flow.
Provided 
$\Lambda_{QCD},M_N\ll P_z \ll \tau^{-1/2}$, the smeared quasi distribution and the light-front PDF 
can be matched through a convolution relation. Alternatives have been suggested, including an exponentiated mass counterterm inspired by heavy quark effective theory and, more recently, RI/MOM and RI$^\prime$ renormalisation schemes.

The chief advantage of our approach is that the gradient flow renders the quasi 
PDF finite in the continuum limit and evades the issues of renormalisation 
at finite lattice spacing. Our approach thus removes the need to consider operator mixing induced by the lattice regulator, since this mixing vanishes as one takes the continuum limit. The resulting continuum matrix elements are 
independent of the choice of lattice action and can be 
matched directly to the corresponding light-front PDFs in the $\overline{MS}$ scheme 
using 
continuum perturbation theory. We are currently undertaking a nonperturbative proof-of-principle calculation to demonstrate the feasibility of our proposal and to better understand systematic uncertainties.

\section*{Acknowledgments}
We thank Carl Carlson, Michael Freid, Anatoly Radyushkin, and Christopher Coleman-Smith
for enlightening discussions during 
the course of this work. K.O.~is supported
by the U.S.~Department of Energy through Grant Number DE-FG02-04ER41302, and 
through contract Number DE-AC05-06OR23177, under which JSA operates the Thomas
Jefferson National Accelerator Facility. C.J.M is supported in part
by the U.S.~Department of Energy through Grant Number DE-FG02-00ER41132.

\bibliography{Lattice2017_38_monahan}

\end{document}